\documentclass[a4paper,10pt]{article}
\date{}
\begin{document}

\title{ Transition from random to ordered fractals in fragmentation of particles in an open system }
\author{{\small M. K. Hassan$^{1,2}$ and J. Kurths$^1$} \\ {\small $~^1$ University of Potsdam, 
Physics Department, Am Neuen Palais, D-14415, Potsdam, Germany} \\
 {\small $~^2$  University of Dhaka, Department of Physics, Theoretical Physics 
Division, Dhaka 1000, Bangladesh}}

\maketitle

\begin{abstract}

\noindent
We consider the fragmentation process with mass loss and discuss self-similar properties 
of the arising structure both in time and space, focusing on dimensional analysis. This 
exhibits a spectrum of mass exponents $\theta$, whose exact numerical values are given
for which $x^{-\theta}$ or $t^{\theta z}$ has the dimension of particle size 
distribution function $\psi(x,t)$, where $z$ is the kinetic exponent. We obtained 
conditions for which the scaling and fragmentation process altogether break down and give 
explicit scaling solution for special case. Finally, we identify a new class of fractals 
ranging from random to non-random and show that the fractal dimension increases with 
increasing order and a transition to strictly self-similar pattern occurs when 
randomness completely ceases.   

\end{abstract}

\noindent
PACS number(s): 05.45.Df, 46.50.+a, 62.20.Mk
\vspace{2mm}

\noindent
The kinetics of the irreversible and sequential breakup of particles occurs in a variety of 
physical processes and has important applications in science and technology. These 
include erosion \cite{kn.has0}, grinding and crushing of solids\cite{kn.has1}, polymer 
degradation and fiber length reduction \cite{kn.has2}, breakup of liquid droplets 
\cite{kn.has3}, etc. to name just a few. In recent years, there has been an increasing 
interest in studying fragmentation, allowing variations that increase the flexibility of 
the theory in matching such conditions of real phenomena such as extension to higher
dimension \cite{kn.has4}, agglomerate erosion \cite{kn.has0}, mass loss \cite{kn.has5}, 
volume change \cite{kn.has6} and fragmentation-annihilation \cite{kn.has7}. The kinetic 
equation approach of fragmentation is described by a linear integro-differential   
equation where mass or size is the only dynamical quantity. This theoretical approach is 
mean field in character since fluctuations are ignored altogether. 

\noindent
In this article, we study the kinetics of fragmentation with continuous mass loss, an 
interesting variant of the classical fragmentation equation introduced by 
Edwards {\it et al.} \cite{kn.has5}. This is relevant in all fragmentation processes 
mentioned earlier where mass loss might occur due to evaporation, oxidation, 
sublimation, dissolution, melting etc., or in Yule-Furry process of cosmic shower theory 
where energy loss occurs due to collision \cite{kn.cos}. Although there exist a series 
of paper devoted to this problem, the explicit scaling solution with exact numerical
exponents, the geometric properties of the arising pattern and the interplay between
fragmentation and mass loss remain unexplored. In addition to obtaining these, we give an 
alternative interpretation of the existing scaling theory by considering dimensional analysis, and we give an 
explicit bound to the exponent of the breakup rate for which the kinetic equation fails. 
A strong motivation for the present investigation came from the desire to know how 
fractal geometry of the resulting object changes with the degree of order, which is 
quantified by the global exponent and is typically known as fractal dimension. 

\noindent
The evolution of the particle size distribution function $\psi(x,t)$ for fragmentation with 
mass loss in one dimension is
\begin {equation}
{{\partial \psi(x,t)}\over{\partial t}}  = -  \psi(x,t)\int_0^x F(y,x-y)dy 
+2\int_x^\infty dy \psi(y,t)F(x,y-x) \nonumber \\ 
 +  {{\partial}\over{\partial x}}(m(x)\psi(x,t)),
\end {equation}
where $F(x,y)$ is the breakup kernel describing the rules and the rate at which a 
particle of size $(x+y)$ breaks into sizes $x$ and $y$. Equation $(1)$ describes a 
process whereby cuts are equivalent to seeds being sown on the fragmenting objects, thus 
producing two new segments. This immediately creates two more new ends belonging to the 
two different, newly created fragments; in doing so, fragments start losing their 
masses immediately from either end until they encounter another seed or become dustlike, 
thereby stopping the loss of their masses. Equation $(1)$ also describes the sequential 
deposition of point-size particles that grow once deposited successfully and stop 
growing upon collision with another point or growing particle. In other words, the 
present model describes nucleation and growth when $\psi(x,t)$ describes the 
gap size distribution of size $x$ at time $t$ or 
how space is covered by growing objects. This occurs in a number 
of natural phenomena, including phase separation, wetting, droplet growth, and growth of 
breath figures. Recently, a variant of the present model was considered in 
\cite{kn.rodgers} in which the deposition of growing rods instead of growing seeds 
was addressed.

\noindent
We consider the breakup kernel to be $F(x,y)=(xy)^\beta (x+y)^{\sigma-1}$, for which the 
breakup rate $a(x)=\int_0^x F(y,x-y)dy=px^{\lambda(\beta,\sigma)}$, where 
$p={{(\Gamma(\beta+1))^2}\over{\Gamma(2 \beta+2)}}$, the homogeneity index 
$\lambda(\beta,\sigma)=2\beta+\sigma$, and $\Re(\beta)>-1$. To comply with the present 
choice of breakup rate $a(x)$, it is essential to consider a similar power law form for 
$m(x)$, hence we choose $m(x) \sim mx^\gamma$, with $m$ a positive real constant. 
However, $a(x)$ being the quantity describing the rate at which particles are 
fragmenting, $x^{-\lambda(\beta,\sigma)}\equiv \tau(x)$ must bear the dimension of time, 
and this put a strong constraint on the exponent $\gamma$. That is, the dimensional 
consistency requires $\gamma= \lambda(\beta,\sigma)+1$. This dimensional consistency has
been ignored in all previous studies \cite{kn.has5}; instead $\gamma < 
\lambda(\beta,\sigma)+1$ was identified as the recession regime and $\gamma > 
\lambda(\beta,\sigma) +1$ as the fragmentation regime. A closer look at the rate 
equation reveals that $x$ and $t$ are inextricably intertwined via the dimensional 
consistency, and hence the system becomes stochastic in nature. So, it is obvious that 
either of the two can be taken to be an independent parameter when the other one is 
expressible in terms of this. For example, if $x$ is chosen to be the independent 
parameter, then using the fact that the dimension of a physical quantity is always 
expressed as power monomial, we can define the dimensionless quantity  
$\xi={{t}\over{x^\alpha}}$. It is obvious that the dimension of the governed quantity 
$\psi(x,t)$ as well can be expressed in terms of $x$ alone, and hence we can define 
another dimensionless quantity $\Pi={{\psi(x,t)}\over{x^{-\theta}}}\sim x^\theta 
\psi(x,x^\alpha\xi)\equiv F(x,\xi)$. Since $\xi$ and $\Pi$ are dimensionless quantities
upon transition from one system of units of measurement to another inside a given class, 
their numerical values must remain unchanged meaning ${{\partial F}\over{\partial 
x}}=0$. This implies that $\Pi$ is independent of $x$ and can be completely expressed in 
terms of $\xi$ alone. Thus we can define $\Pi=\Phi(\xi)$ to enable us to write the 
spatial scaling {\it ansatz} $\psi(x,t) \sim x^{-\theta}\Phi(\xi)$. Had we chosen time
$t$ to be the independent quantity, a similar argument which would lead us to write the
temporal scaling {\it ansatz} $\psi(x,t) \sim t^{\theta z}\phi(\eta)$ with $\eta= 
{{x}\over{t^{-z}}}$. We know that $x^{-\lambda(\beta,\sigma)}$ has the dimension of 
time, therefore $t^{-{{1}\over{\lambda(\beta,\sigma)}}}\equiv \delta(t)$ must have the 
dimension of $x$. This gives us $z={{1}\over{\lambda(\beta,\sigma)}}$, which is known as 
the kinetic exponent since $\delta(t)$ describes the mean or typical cluster size and 
$\alpha=-\lambda(\beta,\sigma)$. Inserting the temporal scaling {\it ansatz} into 
Eq. (1) and requiring scaling to exist would give the same result for kinetic 
exponent $z$. However, dimensional consideration proved to be very instructive and time 
saving, yet rich in physics.

\noindent
Note that the exponent $\theta$ takes the value for which $x^{- \theta}$ and $t^{\theta 
z}$ have a dimension of $\psi(x,t)$, and hence it is called the mass exponent. The existence 
of scaling or a self-similar solution actually means that we can choose self-similar
coordinates ${{\psi}\over{t^{\theta z}}}$ (or ${{\psi}\over{x^{- \theta}}}$) and 
$x/\delta(t)$ (or $t/\tau(x)$) such that their plots for any initial condition collapse 
into one single curve. It is very instructive to note that at $\lambda(\beta,\sigma)=0$,
the governing parameters $x$, $t$ and the governed parameter  $\psi$ all lose their 
dimensional characters. As a result, the system loses its stochastic nature. This means 
that one can no longer define self-similar coordinates at $\lambda(\beta,\sigma)= 0$, 
which is conceptually very important and crucial for scaling to exist, and hence scaling
at $\lambda(\beta,\sigma)\leq 0$ breaks down. Note the inherent properties of the 
fragmentation process, which are the typical or mean cluster size $\delta(t)$, must 
decrease as time proceeds, and the number density must be an increasing function of time. 
In order to be so, the system must be governed by some conservation laws. However, this 
is not true as the homogeneity index $\lambda(\beta,\sigma)<0$, in which case $\delta(t)$ 
becomes an increasing function of time. This simply goes against the principle of 
kinetics of fragmentation. Combining all this, we argue that the system fails to show
scaling not only at $\lambda(\beta,\sigma)<0$, 
as was reported by Cheng and Redner \cite{kn.cheng}, but also 
at $\lambda(\beta,\sigma)=0$. Many authors noticed 
further anomalous behaviour in this regime, e.g., lack of conservation of mass 
\cite{kn.ernst} and an absence of self-averaging \cite{kn.has10}. Due to this anomalous 
behaviour, this regime was termed a transition to shattering \cite{kn.has9}. These 
heurestic arguments actually imply a possible failure to describe physically meaningful 
fragmentation process by the breakup kernel, for which the homogeneity index 
$\lambda(\beta,\sigma) \leq 0$ and the shattering is actually an articulate term for 
this regime. The above discussion is true for Eq. $(1)$ even when the mass loss 
term is absent. Therefore, we shall restrict the rest of the discussion to the regime 
$\lambda(\beta,\sigma) > 0$ with $\Re(\beta)>-1$.

\noindent
We now turn to find the mass exponent $\theta$, which can only be found if the system 
follows some conservation laws. For example, for pure fragmentation ($m=0$), the mass or 
size of the system is a conserved quantity and gives $\theta=2$. Defining the $n^{th}$ 
moment $M_n(t)=\int_0^\infty x^n \psi(x,t)dx$ and combining it with the rate equation 
$(1)$ for the present choice of $F(x,y)$ and $m(x)$ yields 
\begin{equation}
{{dM_n(t)}\over {dt}}=-[{{(\Gamma(\beta+1))^2}\over{\Gamma(2 \beta+2)}}-{{2 
\Gamma(\beta+1)\Gamma(n+\beta+1)}\over{\Gamma(n+2 \beta+2)}}+mn]M_{n+2 \beta + 
\sigma}(t).
\end {equation}
Note that the number density $M_0(t)$ evolves in the same fashion as it would in the absence of the mass loss term. 
This means that particles keep losing their mass in a continuous manner by some mechanisms that do not 
alter the number density. The interesting feature of the above equation is that for $m > 0$, there are infinitely  
many $n=D_f(\beta,m)$ values for which $M_{D_f(\beta,m)}(t)$s are conserved quantities.  
However, for $m=0$, there is only one conserved quantity $M_1(t)$, i.e., size or mass of 
the system, and this does not depend on $\beta$. The dynamics of the system is governed 
by conservation laws and, as a consequence, the system shows scale invariance. These 
conserved quantities in fact are the intrinsic agent responsible for tuning the 
numerical value of the mass exponent $\theta$, and introduce universality to the process.
However, at $\lambda \leq 0$ there is no evidence that there exists any conservation law 
for which the mean cluster size can be a decreasing function of time while the mean 
number density can behave in the opposite way. We can find the numerical value of 
$D_f(\beta,m)$ by searching for the positive and real root of the equation obtained by 
setting the term in the bracket of the Eq. (2) equal to zero, which is polynomial 
in $n$ of a degree determined by the $\beta$ value. Substituting the temporal scaling {\it 
ansatz} into the definition of $M_n(t)$ gives $M_n(t)\sim 
t^{-(n-(\theta-1))z}\int_0^\infty \eta^n \phi(\eta)d \eta$, and demanding 
$M_{D_f}(\beta,m)$ be a conserved quantity immediately gives $\theta=(1+D_f(\beta,m))$, 
which clearly depends on $\beta$ and $m$ if and only if $m>0$. Owing to the random nature of 
the process and due to the presence of mass loss term, it is clear that when the process 
continues {\it ad infinitum}, it creates a distribution of points (dust) along a line at
an extreme late stage that is different from any known set  \cite{kn.has11,kn.has12}.  
To measure the size of the set created in the long-time limit, we define a line segment 
$\delta(t)={{M_1(t)}\over{M_0(t)}}\simeq t^{-{{1}\over{\lambda(\beta,\sigma)}}}$, which 
is typical cluster size. We can count the number of such segments needed to cover the 
set, and in the limit $\delta \longrightarrow 0$ (i.e., $ t \longrightarrow \infty$), the 
number $N(\delta)$ will simply  measure the set and appear to scale as $N(\delta) \sim 
\delta^{-D_f(\beta,m)}$. The exponent $D_f(\beta,m)$ is known as the 
Hausdorff-Besicovitch  \cite{kn.has12} dimension or the fractal dimension of the arising 
pattern.

\noindent
To get a physical picture of the role played by $m$, we set $\beta=0$ for the time being, 
for which the polynomial equation becomes quadratic in $n$ and the real positive root is 
$D_f(m)=-{{1}\over{2}}(1+1/m)+{{1}\over{2}}\sqrt{(1+1/m)^2+4/m}$
when the second root is $D=-(D_f(m)+1+1/m)$. Therefore, the exponent $\theta$ is also 
function of $m$. The expression for $D_f(m)$ reveals that as value of $m$ increases, the 
fractal dimension decreases very sharply and in the limit $m \longrightarrow \infty$,  
$D_f(m)\longrightarrow 0$. This means that as $m$ increases, the size of the 
corresponding arising set decreases sharply due to fast disappearance of its member, 
whereas, as $m \longrightarrow 0$, $D_f(m) \longrightarrow 1$, that is, we recover the 
full set (pure fragmentation) that describes a line. On the other hand, had we kept $m$ 
fixed and let $p$ decrease, the effect would have been the same as we observed for 
increasing $m$ with $p=1$ (i.e. $\beta =0$). Thus, it is the ratio between $m$ and $p$ 
that matters rather rather than their individual increases or decreases. To give further 
physical picture of what these results mean, we define mass length relation for the 
object as $M_0 \sim \delta^{D_f(m)}$ and $M_e \sim \delta^d$ for the space where the 
object is being embedded, where $d$ describes the Euclidean space. The density of the 
property of the object $\rho$ then scales as $\rho \sim \delta^{D_f(m)-d}$. It is thus 
clear that for a given class of set created by a specific rule, when $D_f(m)$ decreases 
it means that it is increasingly moving away from $d$ and hence  more and more members 
are removed from the full set. This in turn creates increasingly ramified or stringy 
objects, since $D_f(m)=d$ describes the compact object with uniform density. Therefore, this 
shows that increasing $m/p$ ratio means that the mass loss process becomes stronger than the 
fragmentation process and vice versa.

\noindent
We now attempt to find the spatial scaling solution  for $\Phi(\xi)$. Note that the 
dimension of the arising pattern is independent of $\sigma$ and consequently independent 
of how fast or slow the system performs the process. Therefore, we can set $\sigma=1$ without 
risk of missing any physics, but it certainly simplifies our calculation. Substituting
the spatial scaling {\it ansatz} into the rate equation (1) for $F(x,y)=1$ and 
$m(x)=mx^2$ and differentiating it with respect to $\xi$, transforms the partial 
integro-differential equation into an ordinary differential equation, 
\begin{equation}
\xi(1-m \xi)\Phi^{\prime \prime}(\xi) 
+\{(1-\theta)-\xi[2m(2-\theta)-1]\}\Phi^{\prime}(\xi)-[m(2-\theta)(1-\theta)
-(3-\theta)]\Phi(\xi)=0.
\end{equation}
For $m=1$, this is hypergeometric differential equation \cite{kn.hass} whose only 
physically acceptable linearly independent solutions are $~_2F_1(1,-(1+2D_f);-D_f;\xi)$ 
and $\xi^{(1+D_f)}~_2F_1(2+D_f,-D_f;2+D_f;\xi)$, where $D_f=0.414213$. From these exact 
solutions for spatial scaling function, we can obtain the asymptotic temporal scaling 
function $\phi(\xi) \sim e^{-D_f \xi}$ which satisfies the condition 
$\phi(\xi)\longrightarrow 0$ as $\xi \longrightarrow \infty$.

\noindent
We now attempt to see the role of $\beta$ on the system. To judge its role, it is clear 
from the previous discussion that we ought to give equal weight to all the terms in 
Eq. (1) so that each of them can compete on an equal footing. This can be done if 
we set $m=p={{(\Gamma(\beta+1))^2}\over{\Gamma(2\beta+2)}}$ so that the relative 
strength  between fragmentation and the mass loss process stays the same as the value of $\beta$  
increases. This is a very crucial point to be emphasized. We can obtain the fractal 
dimension for different values of $\beta$, which is simply the real positive root of the 
polynomial equation in $n$ of degree $\beta$. A detailed survey reveals that the fractal 
dimension increases monotonically with increasing $\beta$. To find the fractal dimension
in the limit $\beta \longrightarrow \infty$, we can use the Stirling's approximation in 
the polynomial equation to obtain $\ln[n+1]=(1-n)\ln[2]$ when $n=0.4569997$ solves this 
equation. In order to give a physical picture of the role of $\beta$ in the limit $\beta 
\longrightarrow \infty$, we consider the following model: $F(x,y)=(x+y)^\gamma 
\delta(x-y)$. This model describes that cuts are only allowed to be in the middle in 
order to produce two fragments of equal size at each time event. This makes 
$a(x)={{1}\over{2}}x^\gamma$, so we need to choose $m(x)={{1}\over{2}}x^{\gamma+1}$, 
($m={{1}\over{2}}$ gives the same weight as for the fragmentation process). Then the 
rate equation for $M_n(t)$ becomes $M_n^\prime(t)=-[(n+1)/2-2^{-n}]M_{n+\gamma}(t)$.
As before, we set the numerical factor on the right-hand side of this equation equal to 
zero and then take the natural logarithm on both sides to obtain the $n$ value for which $M_n(t)$ 
is time independent. In doing so, we arrive at the same functional equation for $n$ as 
we found for $\beta \longrightarrow \infty$. This shows that the kernel 
$F(x,y)=(xy)^\beta (x+y)^{\sigma-1}$ behaves exactly in the same fashion as for 
$F(x,y)=(x+y)^\gamma \delta(x-y)$. We thus find that in the limit $\beta \longrightarrow 
\infty$, the resulting distribution of points is a set with fractal dimension 
$D_f=0.4569997$, which is a strictly self-similar fractal as randomness is seized by
dividing fragments into equal pieces. We are now in a position to give a physical 
picture of the role played by $\beta$. First of all, the process with $\beta=0$ 
that describes the frequency curve of placing cuts about the size of the fragmenting 
particles is Poissonian in nature. Consequently, the system enjoys the maximum randomness 
and the corresponding fractal dimension is $D_f=0.414213$. For $\beta >0$, the 
frequency curve of placing cuts about the size of the fragmenting particles is Gaussian 
in nature, meaning that as the value of $\beta$ increases, the particles are increasingly more likely to 
break in the middle than on either end. That is, as $\beta$ increases, the variance 
decreases in such a manner that in the limit $\beta \longrightarrow \infty$, the variance 
of the frequency curve becomes infinitely narrow, meaning a $\delta$-function distribution 
for which the fragments are broken into two equal pieces. Therefore, there is a spectrum of fractal 
dimensions between $\beta \rightarrow 0$ when $D_f=0.414213$ and $\beta \longrightarrow 
\infty$ when $D_f=0.4569997$. A detailed numerical survey that we do not present here, 
confirms that the fractal dimension increases monotonically with $\beta$ and reaches a 
constant value when $\beta \longrightarrow \infty$. The previous density-dimension 
relation implies that increasing $\beta$ vis-a-vis increasing order also means that the 
system loses less and less mass, and this happens despite the fact that now the  
${{m}\over{p}}$ ratio stays the same. This shows that there exists an interplay between 
fragmentation and the mass loss process that can be tuned either by changing the ratio of 
$m$ and $p$, which is obvious of course, or by changing the degree of order alone, which 
is indeed a nontrivial result.

\noindent
In summary, we have identified a new set with a wide range of subsets produced by tuning 
the degree of randomness only. The process starts with an initiator of unit interval 
$[0.1]$, and the generator divides the interval into two pieces, and deleting some parts 
from either side of both pieces at each time step. The amount of parts to be 
deleted is determined by the parameter that controls the intensity of randomness. We 
quantified the size of the resulting set obtained in this way by fractal dimension and 
showed that the fractal dimension increases with increasing order and reaches its 
maximum value when the pattern described by the set is perfectly ordered, which is 
contrary to some recently found results \cite{kn.has13}. We also discussed the scaling 
theory of the process, emphasizing dimensional analysis, and we showed that the shattering 
is in fact is an articulate term whereby the equation fails to describe physically 
meaningful fragmentation process. We have also shown that the interplay between 
fragmentation and mass-loss arises not only from the ratio of their strengths determined 
by their respective numerical coefficients, but also by the degree of order. We give 
exact numerical value of the mass exponent, which has never been reported, and we obtained 
the explicit scaling function for special case of interest. Finally, we argue on the 
basis of our findings that fractal dimension, degree of order and the extent of 
ramifications of the arising pattern are interconnected. 

\noindent
The author is grateful to R. M. Ziff for sending valuable comments and acknowledges
inspiring correspondence with P. L. Krapivsky. MKH  acknowledges the Alexander von
Humboldt Foundation for granting the fellowship

\end{document}